\documentclass[12pt]{iopart}

\usepackage{iopams}  
\usepackage{graphicx}
\usepackage{booktabs} 
\usepackage{subcaption}
\usepackage{epstopdf}
\usepackage{caption}
\usepackage{placeins}
\usepackage{float}
\usepackage{xcolor}
\usepackage{everypage}

\begin{document}

\title{Interpretable Classification of Levantine Ceramic Thin Sections via Neural Networks}

\author{
Sara Capriotti$^1$, 
Alessio Devoto$^2$, 
Simone Scardapane$^3$, 
Silvano Mignardi$^1$, 
Laura Medeghini$^1$
}

\address{$^1$ Department of Earth Sciences, Sapienza University of Rome, P.le Aldo Moro 5, 00185 Rome, Italy}
\address{$^2$ Department of Computer, Control and Management Engineering, Sapienza University of Rome, Via Ariosto 25, 00185 Rome, Italy}
\address{$^3$ Department of Information Engineering, Electronics and Telecommunications, Sapienza University of Rome, Via Eudossiana 18, 00184 Rome, Italy}

\ead{sara.capriotti@uniroma1.it}

\begin{indented}
\item[] Corresponding author: Sara Capriotti
\end{indented}

\begin{abstract}
\raggedright
Classification of ceramic thin sections is fundamental for understanding ancient pottery production techniques, provenance, and trade networks. Although effective, traditional petrographic analysis is time-consuming. This study explores the application of deep learning models, specifically Convolutional Neural Networks (CNNs) and Vision Transformers (ViTs), as complementary tools to support the classification of Levantine ceramics based on their petrographic \textit{fabrics}. A dataset of 1,424 thin section images from 178 ceramic samples belonging to several archaeological sites across the Levantine area, mostly from the Bronze Age, with few samples dating to the Iron Age, was used to train and evaluate these models. The results demonstrate that transfer learning significantly improves classification performance, with a ResNet18 model achieving 92.11\% accuracy and a ViT reaching 88.34\%. Explainability techniques, including Guided Grad-CAM and attention maps, were applied to interpret and visualize the models' decisions, revealing that both CNNs and ViTs successfully focus on key mineralogical features for the classification of the samples into their respective petrographic \textit{fabrics}. These findings highlight the potential of explainable AI in archaeometric studies, providing a reproducible and efficient methodology for ceramic analysis while maintaining transparency in model decision-making.
\end{abstract}

%
\vspace{2pc}
\submitto{Machine Learning: Science and Technology}
%
%
%
 
\section{Introduction}

The archaeological context of the Levantine region is both particularly significant and complex, as numerous archaeological sites emerged during the Early Bronze Age, a period characterized by urbanization, trade networks, and political complexity \cite{Steiner2014, Greenberg2019}. During this phase, the first urban societies in the region took shape. The process of urbanization was marked by the aggregation of local settlements, the formation of regional centers, and the fortification of villages, reflecting a shift toward more centralized form of organization. These transformations included the simplification of material culture and the reorganization of craft production \cite{Greenberg2011}. One example of this process is Tell el-Far’ah (North), with strong fortifications and standardized ceramic production suggesting the presence of large workshops and inter-community exchange. While urbanization began earlier in the southern areas of the Levantine region, northern parts followed this path a bit later with key sites like Ebla showing signs of complex administration and organized settlement planning \cite{Weiss2013, Chesson2003}.Therefore, over the past decades, there has been a growing interest in studying this region to understand and reconstruct the changes and developments that have been observed in various archaeological sites. The study of ancient ceramics plays a central role in understanding the social and cultural identity of the region from the Bronze Age, as it provides information on social, cultural, and technological development as well as the evolution of ancient civilizations. Indeed, an archaeometric characterization of ancient ceramic materials can define the nature and provenance of raw materials used in production that help to define trade routes and interaction among past societies \cite{Quinn2013, Orton2013, Yu2018}. 

Currently, petrographic analysis of ceramic thin sections is one of the most effective and widely used methods for archaeometric characterization and reconstruction of the technological and material features of ceramics. The petrographic analysis also includes the identification of \textit{fabrics}, groups of thin sections sharing similar characteristics, and representative of the \textit{chaîne opératoire}, i.e. the actions and technological choices that led to the formation of the final product \cite{Whitbread1986, Whitbread1995}. Therefore, the grouping of thin sections in \textit{fabrics} has been widely applied to define the different “recipes” used in ceramic production and to distinguish between local or imported wares \cite{Whitbread2017}. Specifically, petrographic analyses can help in the recognition of specific ware types as locals and in identifying patterns of exchange of pottery within the region \cite{Goren1996, Greenberg1996, Goren2010, DAndrea2015}. In other cases, they have contributed to the identification of relationships between ancient production centers and their surrounding geological settings \cite{Maritan2005, Badreshany2020, Tumolo2023}. However, reconstructing ancient exchanges and defining trade profiles remain challenging due to the still limited petrographic studies on large ceramic assemblages, resulting in a lack of comparative data useful for identifying potential relationships and interactions. Moreover, although minero-petrographic characterization is a well-established and effective method for studying pottery assemblages, it is time-consuming and challenging to compare results across different archaeological contexts. This is partly due to the dispersion of petrographic data across various sources such as specialist journals, monographs, conference proceedings, and excavation reports.

In recent decades, there has been a growing interest in applying Machine Learning (ML) and/or Deep Learning (DL) techniques to archaeology in a variety of ways, including the identification of archaeological sites and structures \cite{caspari2019, Guyot2021, Trier2019}, and the classification or recognition of archaeological artifacts. Recent applications have involved the use of neural networks for the prediction of morphological features of lithic artifacts even when these are fragmented or incomplete \cite{troiano2024, Nobile2024}. Other applications include the use of Laser-Induced Breakdown Spectroscopy (LIBS) combined with neural networks to attribute archaeological bone remains to specific individuals \cite{Panagiotis2021}. Additionally, Lithic Use-Wear Analysis (LUWA) using microscopic images has been employed to study the working traces on lithic artifacts \cite{zhang2024}. Concerning the study of ceramics, automated methodologies have become powerful tools for resolving classification tasks, significantly contributing to the recognition of specific compositional, technological, or stylistic patterns \cite{Bickler2021, Pawlowicz2021}. Today, several algorithms have been tested for the classification of pottery based on the study of their typology, decoration, and shape \cite{Horr2014, Navarro2021}. ML models have also been applied in provenance analysis, where ceramic samples have been classified through the elaboration of geochemical data \cite{Qi2022, Ruschioni2023, Anglisano2022}. Among these approaches, Artificial Neural Networks (ANNs) have been employed to determine ceramic provenance using LIBS spectra \cite{Ramil2008}. ANNs have further been used for the analysis of ceramic thin sections with the aim of facilitating the \textit{fabric} classification, in combination with image analysis to identify inclusions and pores in thin sections by working with parameters such as color, shape, size, percentage of inclusions and porosity \cite{Aprile2014}. Recent studies focused on the development of a DL model based on Convolutional Neural Networks (CNNs) to classify ceramic thin sections into their respective petrographic \textit{fabrics} \cite{Lyons2021, Lyons2022}. 

CNNs are specifically designed for image pattern recognition. They consist of a series of layers among which the convolutional ones play a fundamental role, allowing the detection of characteristic patterns meaningful for accurate classification \cite{LeCun2015}. 

Recently, Transformers-based models have emerged as a significant alternative to CNNs. Vision Transformers \cite{dosovitskiy2021an} (ViTs) operate using multi-head self-attention instead of convolution and process input images by dividing them into patches. This approach captures both local and global dependencies, facilitating the identification of relationships across the entire image \cite{Raghu2021}. Yang et al. have explored the combination of channel and self-attention mechanisms within CNN architectures to improve their discriminative power in the classification of Ming-Qing ceramics \cite{Yang2025}.

Despite being extremely accurate, modern DL methods are highly complex, often consisting of hundreds of layers and parameters, making it difficult to determine which features the model relies on for classification. As a consequence, predictions are usually difficult to explain. For this reason, such models are often considered ‘black boxes’, as they do not provide any direct explanation or insights about their predictions \cite{Castelvecchi2016}. In order to tackle this limitation, Explainable Artificial Intelligence (XAI) emerged as a field aimed at studying the interpretability of ML models \cite{Zhong2022, Arrieta2020}. Understanding what the model has learned allows for further validation of its accuracy, also ensuring that it is based on the correct variables during the training phase, allowing the identification of potential limitations and building trust in the model's performance \cite{Oviedo2022, Montavon2018}. In the context of XAI, saliency maps and attention maps are visual tools that help to understand a model's decision \cite{genovesemixture}. To the best of our knowledge, neither ViTs nor explainability techniques have been applied to the ceramic petrographic \textit{fabric} classification. The only similar study applied Grad-CAM (Gradient-based Class Activation Maps) to visualize the key features used by a CNN model for petrographic classification, although explainability was not its main focus \cite{Lyons2022}. Other related works have employed Grad-CAM to interpret CNN predictions  for the typological and/or chronological classification of ceramic sherds and vessels \cite{Pawlowicz2021, ElHajj2023,Yang2025}.

The present research focuses on the application of automated methodologies for the classification of Levantine ceramics into their petrographic \textit{fabrics} using deep learning models, specifically Convolutional Neural Networks (CNNs) and Vision Transformers (ViTs). The aim is to investigate the potential of these models in the field of petrographic studies of pottery. The performance of these models is compared to evaluate their effectiveness in classification tasks. Furthermore, the research provides a partial interpretation of CNNs and ViTs using several forms of visual explanations. The objective is a deeper understanding of the classification process by visualizing the distinctive features that most influenced the models' predictions.

\section{Materials and Methods}
\label{sec:materials}

\begin{figure}[ht]
    \centering
    \includegraphics[width=0.75\textwidth]{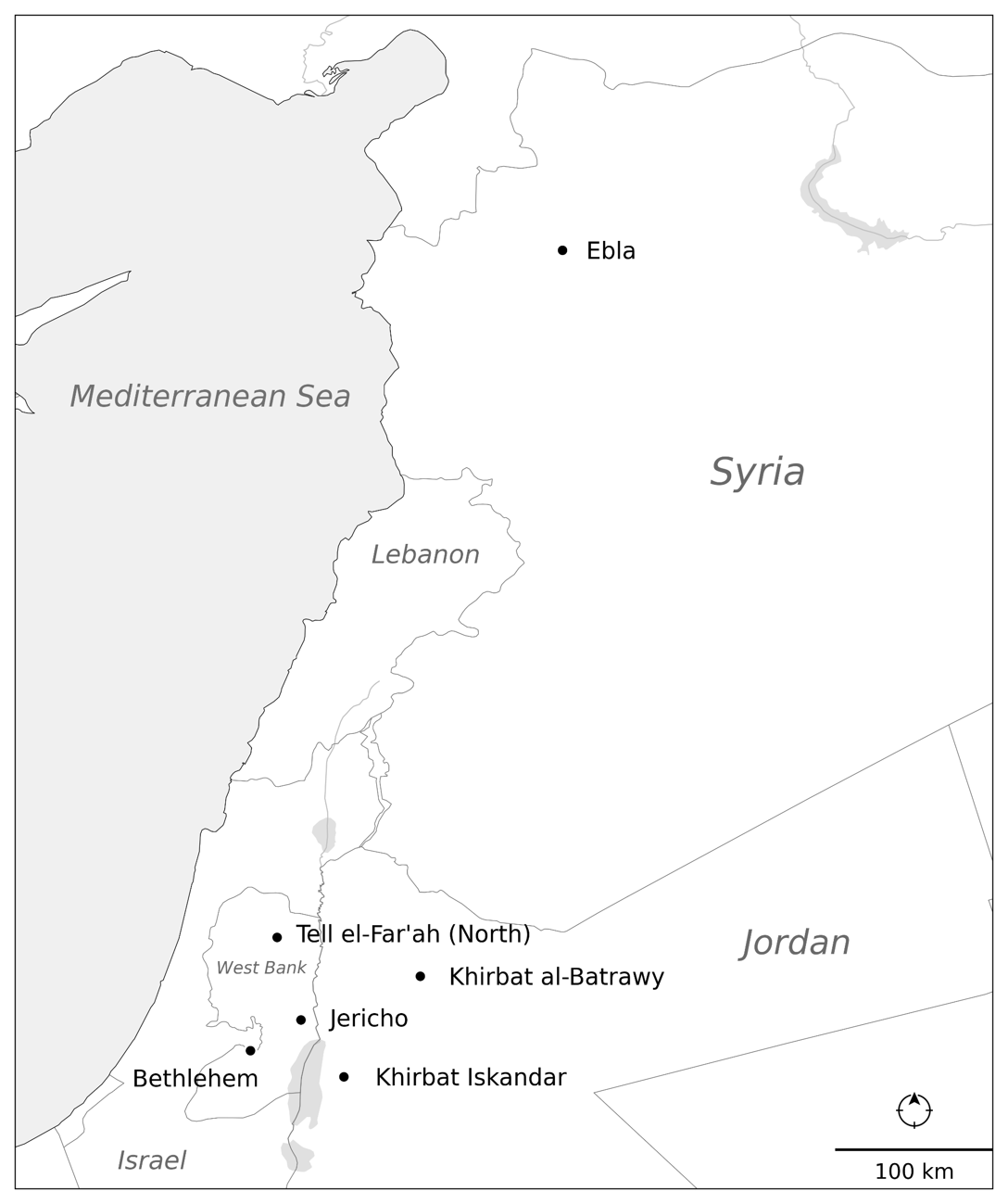}
    \caption{Map of the Levant showing the selected archaeological sites.}
    \label{figure1}
\end{figure}

The starting point for the implementation of the DL models consisted of a phase of selection and acquisition of a large number of ceramic thin section images, which have been used as training and testing datasets. A total of 178 ceramic samples were selected, primarly dating to the Bronze Age, with a smaller number from the Iron Age. These samples come from six different archaeological sites across the Levantine region: Bethlehem (West Bank), Tell el-Far'ah North (West Bank), Khirbat Iskandar (Jordan), Khirbat al-Batrawy (Jordan), Ebla (Syria) and Jericho (West Bank) as shown in Fig. \ref{figure1} (see Table \ref{table1} for more detailed information on the samples). While most of the sites are located in the southern part of the region, the inclusion of Elba provides important comparative data from the north. All samples have already been described and classified in terms of their respective petrographic \textit{fabrics}, as well as their archaeological context in previous papers \cite {Medeghini2019, Medeghini2016, Botticelli2020, Botticelli2022, Sala2023, Medeghini2016KB, Ballirano2014, Medeghini2017}. The selection aimed to create a dataset representative of the Levantine pottery production profile, according to manufacturing practices, technological choices, and regional variability.

\renewcommand{\thetable}{\arabic{table}}
\begin{table}
    \caption{\label{table1}Summary of the selected ceramic thin sections.}
    \begin{indented}
    \lineup
    \item[]\begin{tabular}{@{}l l c}
        \br
        {\textbf{Archaeological sites}} & {\textbf{Age}} & {\textbf{N. of samples}} \\
        \mr
        Tell el-Far’ah (North) & Early Bronze I - II & 55 \\
        Khirbat al-Batrawy & Early Bronze III - IV & 40 \\
        Khirbat Iskandar & Early Bronze IV & 35 \\
        Bethlehem & Early Bronze IV - Iron Age IB-II & 23 \\
        Ebla & Early Bronze IV - Middle Bronze I & 18 \\
        Jericho & Early Bronze & 7 \\
        \br
    \end{tabular}
    \end{indented}
\end{table}

\begin{figure}[ht]
    \centering
    \includegraphics[width=1\textwidth]{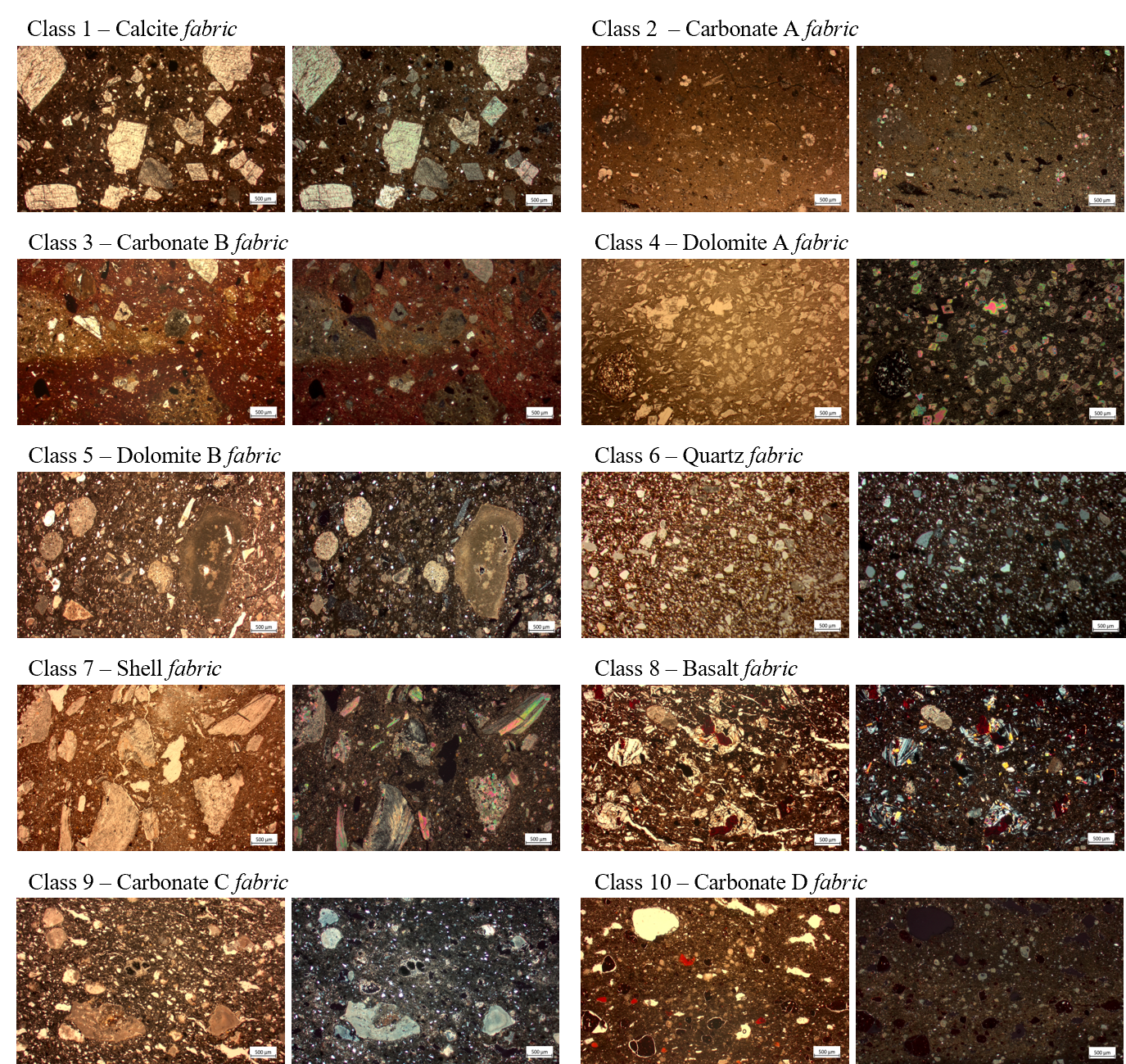} 
    \caption {Thin section images of representative samples of the ten petrographic \textit{fabrics} (2.5X, both in PPL and XPL respectively). Class 1 (sample T53): Characterized by the predominance of calcite inclusions; Class 2 (sample TF6): Predominance of fragments of sedimentary calcareous rocks, presence of microfossils in the matrix; Class 3 (sample TFN39): Fragments of sedimentary calcareous rocks, presence of quartz in the matrix; Class 4 (sample B1/14): Predominance of dolomite crystal inclusions; Class 5 (sample K7): Presence of dolomite, calcareous and siliceous rock fragments as dominant inclusions; Class 6 (sample TFMcamp1): Presence of predominant quartz inclusions; Class 7 (sample K12): Predominant presence of elongated bivalve shells and calcareous rock fragments; Class 8 (sample E467/3): Dominant presence of volcanic rock fragments identified as basalt; Class 9 (sample K11): Characterized by the dominant presence of calcareous and siliceous rock fragments; Class 10 (sample KB314/113): Predominance of calcareous fragments and iron oxide nodules. Complete description of these \textit{fabrics} in \cite{Medeghini2019, Medeghini2016, Botticelli2020, Botticelli2022, Sala2023, Medeghini2016KB, Ballirano2014, Medeghini2017}.}
    \label{figure2}
\end{figure}

Acquisition of thin section images was performed using the Axiocam 208 ZEISS camera connected to a ZEISS D-7082 Oberkochen petrographic microscope. For each ceramic section, we took multiple images under plane-polarized light (PPL) and cross-polarized light (XPL) at magnifications of 2.5X and 10X, resulting in a total of eight images per section to ensure the capture of important microstructural and mineralogical features. All images were compared and grouped into ten petrographic \textit{fabrics}, taking into account the results of the previous works. Samples from the different \textit{fabrics} are presented in Fig. \ref{figure2}. The final dataset, consisting of 1,424 images grouped in 10 classes, is randomly split into training and testing sets with a ratio of 80:20. The larger portion was used for the training phase, while the smaller one was used to test the models on new data. 

\subsection{Implementation of CNN and ViT models}
\label{subsec:models_exp}

We select two types of deep learning models for this study: A CNN using the ResNet18 \cite{resnet} architecture, and the more recent Vision Transformer \cite{dosovitskiy2021an}. CNNs are neural networks that process images through multiple convolutional layers, each detecting increasingly complex features. ResNet18 consists of 18 layers organized in 5 stages: an initial 7×7 convolutional layer followed by 8 blocks, each containing two 3×3 convolutional layers with batch normalization and ReLU activation. Each block includes a skip connection that adds the input to the block's output, helping to prevent vanishing gradients. The network concludes with an average pooling layer and a fully connected layer for classification. 

Vision Transformers represent a fundamentally different approach: they first divide input images into fixed-size patches (16×16 pixels in our case), linearly embed each patch, add positional embeddings, and process them through a series of transformer encoder blocks. Each encoder block contains a multi-head self-attention layer followed by a multi-layer perceptron (MLP). Our ViT implementation uses 12 transformer encoder blocks, each with 12 attention heads and a hidden dimension of 384.

We train our models to predict the class of a given image, minimizing a cross entropy loss function. To evaluate classification performance, we train the models and subsequently compare their performance on the held-out test dataset. 

For both models, we use a transfer learning approach, fine-tuning pre-trained versions on the large ImageNet dataset \cite{Deng2009}. This approach leverages the pre-training on ImageNet, a large dataset of 1.2 million images across 1,000 classes. We retain the pre-trained weights of the backbone layers, that have already learned to extract general visual features like edges, textures, and shapes, and only retrain the final classification layers to adapt to our specific classes. This transfer learning strategy allows the model to build upon previously learned features, making it more effective when working with limited data.

As part of our ablation study, we also evaluate a from-scratch approach, where all model parameters are randomly initialized and trained exclusively on our dataset.

We trained the ResNet-18 models for 60 epochs (from scratch) and 25 epochs (pre-trained) respectively. Both implementations used the AdamW optimizer \cite{kingma2014adammethodstochasticoptimization} with learning rates and weight decay values of 3e-4. An adaptive learning rate scheduler was incorporated to dynamically adjust the learning rate throughout the training process.
For the ViT models, training extended across 125 epochs (from scratch) and 60 epochs (pre-trained) using AdamW with a learning rate of 1e-4 and 1e-5 respectively, and a weight decay of 1e-4. To enhance model generalization and effectively expand our training dataset, we implemented various data augmentation techniques \cite{yang2023imagedataaugmentationdeep} including random axis flipping, color jittering, and random cropping. These augmentations were applied consistently during the training phase for both models to mitigate overfitting. We provide further details concerning the training hyper-parameters in \ref{appendixA-hyperparams}.

\section{Results}
\label{sec:results}

The mean and standard deviation of the metrics for CNNs and ViTs are summarized in Table \ref{table2}. The reported values correspond to the average performance of the models initialized three times with different random seeds. For reference, we also show the results of the models trained from scratch without a pre-trained ImageNet checkpoint. As expected, the transfer learning approach led to a significantly higher performance. The ResNet18 model trained from scratch achieved an accuracy of 76.33 ± 0.58\% after 60 epochs with a precision of 78.10 ± 0.73\%, a result consistent with the small size of the dataset. In contrast, the pre-trained ResNet18, initialized with ImageNet weights, reached an accuracy of 92.11 ± 0.44\% and a precision of 92.05 ± 1.35\% in 25 epochs. ViTs are generally more effective for large datasets; as expected, the model trained from scratch underperformed, with an accuracy of 61.01 ± 0.33\% and a precision of 58.74 ± 1.21\% after 125 epochs. However, the pre-trained implementation performed well despite the small dataset, achieving an accuracy of 88.34 ± 0.29\% and a precision of 88.44 ± 0.50\% in 60 epochs, highlighting the effectiveness of transfer learning even with limited data.

\begin{figure}[!ht]
    \centering
        \includegraphics[width=0.9\textwidth]{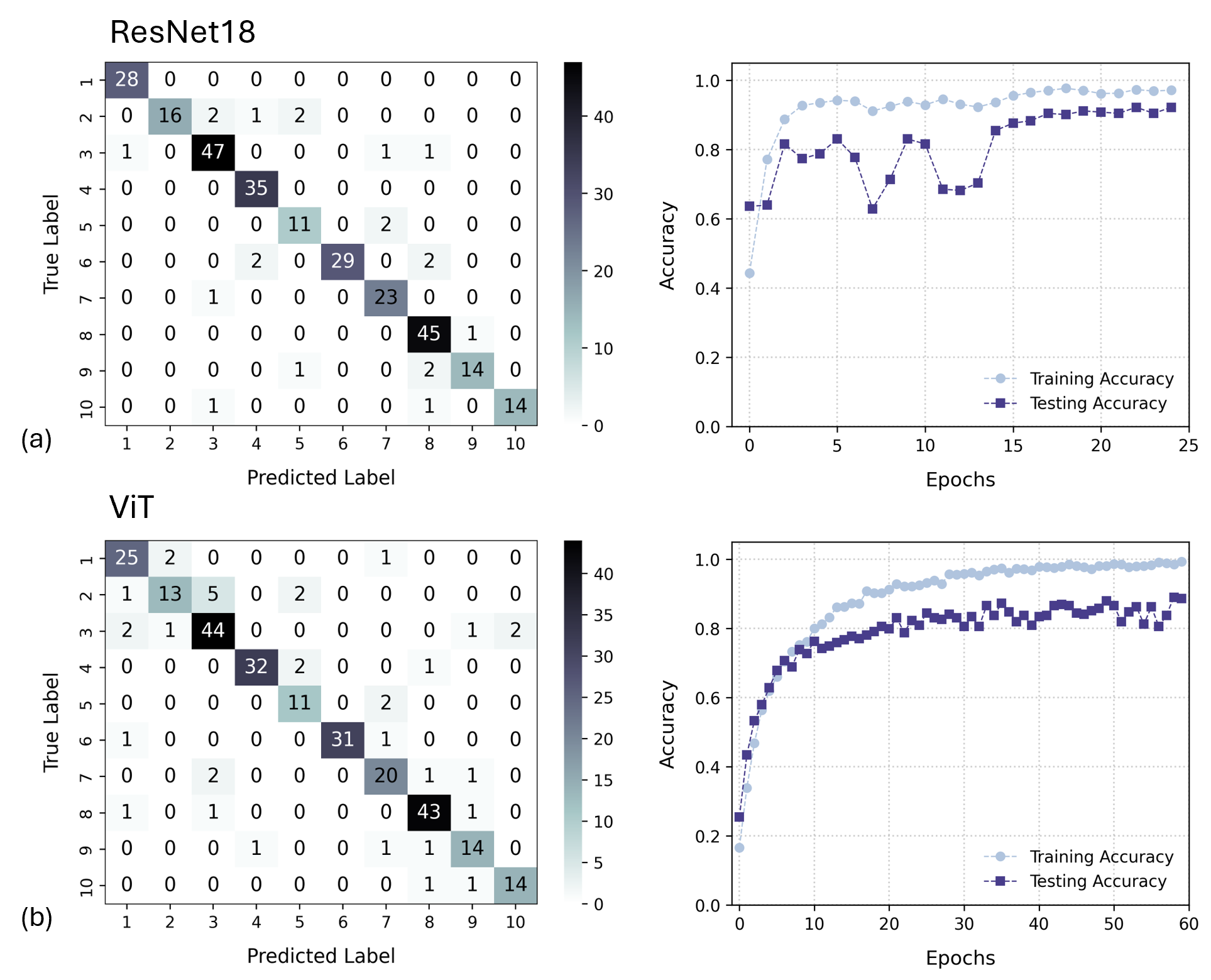}
    \caption{Classification performance of ResNet18 and Vision Transformer (ViT). (a) Confusion matrix and accuracy trends for ResNet18. (b) Confusion matrix and accuracy trends for ViT.}
    \label{figure3}
\end{figure}

Fig. \ref{figure3}a presents the confusion matrix from the best run of the pre-trained ResNet18 model, showing the correlation between predicted and actual values and highlighting the number of correctly classified samples. The adjacent plot illustrates the evolution of training and testing accuracy over the epochs, allowing us to evaluate the learning progress of the model and identify potential overfitting or underfitting issues. Similarly Fig. \ref{figure3}b visualizes the best run results of the ViT model, presenting its confusion matrix along with the training and testing accuracy trends over a total of 60 epochs. It is important to note that the accuracy on the test set was monitored for visualization purposes only. The final evaluation of each model was based on its performance on the test set after training was completed.

\renewcommand{\thetable}{\arabic{table}}
\begin{table}
    \caption{\label{table2}Mean and standard deviation of the accuracy, precision, recall (True positive rate) and F1-score (harmonic mean of precision and recall) metrics of the CNN and ViT.}
    \begin{indented}
    \lineup
    \item[]\resizebox{0.8\textwidth}{!}{%
    \begin{tabular}{@{}l l l l l}
        \br
        {\scriptsize\textbf{Model Metrics}} & {\scriptsize\textbf{ResNet18 (Scratch)}} & {\scriptsize\textbf{Pre-trained ResNet18}} & {\scriptsize\textbf{ViT (Scratch)}} & {\scriptsize\textbf{Pre-trained ViT}} \\  
        \mr
        Accuracy & 76.33 ± 0.58\% & 92.11 ± 0.44\% & 61.01 ± 0.33\% & 88.34 ± 0.29\% \\  
        Precision & 78.10 ± 0.73\% & 92.05 ± 1.35\% & 58.74 ± 1.21\% & 88.44 ± 0.50\% \\
        Recall & 72.16 ± 1.08\% & 90.15 ± 1.58\% & 59.93 ± 0.60\% & 86.60 ± 0.83\% \\
        F1-score & 73.67 ± 0.71\% & 90.68 ± 1.54\% & 58.03 ± 0.47\% & 86.84 ± 0.70\%\\
        \br
    \end{tabular}
    }
    \end{indented}
\end{table}

\subsection{Analysis of misclassified samples}

\begin{figure}[ht!]
    \centering
        \includegraphics[width=0.9\textwidth]{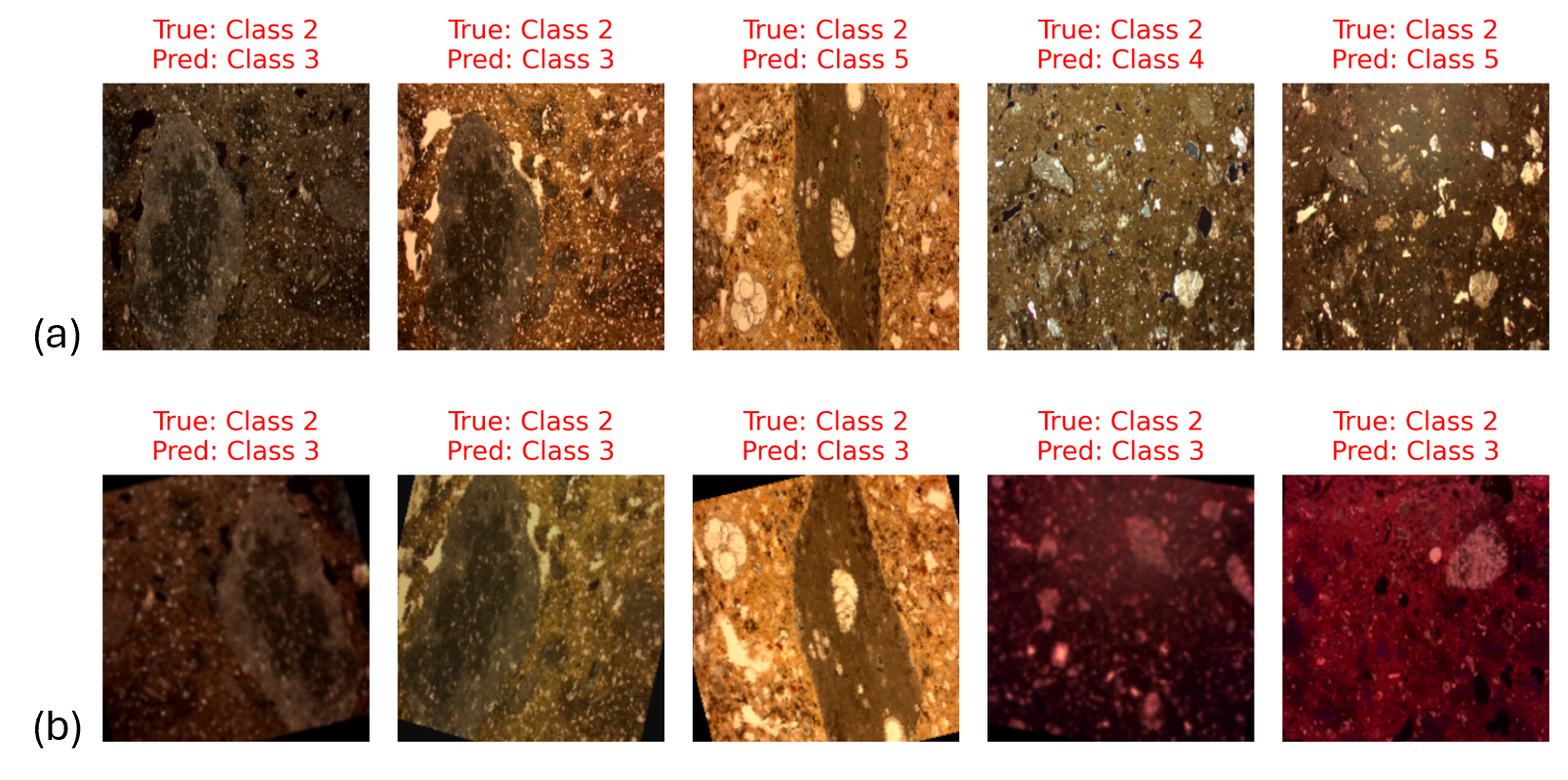}
    \caption{Comparison of ResNet18 and ViT misclassified predictions for class 2, representative of carbonate A \textit{fabric}. (a) Images of ResNet18 misclassified predictions; (b) Images of ViT misclassified predictions.}
    \label{figure4}
\end{figure}

In this section we perform an in-depth analysis of the confusion matrix of the models to verify whether the most difficult samples to assign to specific petrographic \textit{fabrics} correspond to those misclassified by the models and also whether these samples are consistently misclassified, identifying potential biases or problematic samples. From the analysis of the confusion matrix of both the models (Fig. \ref{figure3}), it emerges that class 2 has the highest number of misclassifications with class 3. Class 2 represents the carbonate A \textit{fabric}, characterized by the presence of microfossils in the matrix, while the main inclusions are carbonate rock fragments, making it very similar to the carbonate B \textit{fabric} (class 3). The main difference between them is that carbonate B \textit{fabric} contains an abundant presence of quartz instead of microfossils \cite{Medeghini2019}. Observing the misclassifications of the ResNet18 and ViT models (Fig. \ref{figure4}a and b), we note that the first three images of sample J3 are misclassified in both models, mainly being confused with class 3. It is interesting to highlight that this sample was particularly difficult to classify, as its matrix contains both microfossils and dispersed quartz. This combination of characteristics places it in an intermediate zone between carbonate A and carbonate B \textit{fabrics}, making its classification more complex. These observations suggest that both the models can struggle with classes that share similar mineralogical characteristics, particularly when a sample exhibits features of multiple petrographic \textit{fabrics}. This kind of ambiguity also makes classification difficult for human experts, suggesting that some misclassifications may reflect the complexity of the samples more than limitations in model performance.

\section{Explainability analysis}
\label{sec:discussion}

\subsection{Explainability Techniques}

Given that our study involves two distinct neural network architectures, ResNet18 and ViT, we employ different explainability techniques suited to their respective structures.

For ResNet18, we apply the Guided Gradient-weighted Class Activation Mapping (Guided Grad-CAM) \cite{Selvaraju_2019}, a widely used interpretability technique for convolutional neural networks. Grad-CAM generates a coarse localization map by computing the gradient of the output with respect to the feature maps in the last convolutional layer, highlighting the most influential regions in the input image for the network's decision.

For the Vision Transformer, we employ attention map visualization to interpret model decisions. Unlike CNNs, which rely on local receptive fields, ViTs use self-attention mechanisms to establish long-range dependencies between tokens \cite{Kashefi2023}. Each ViT layer consists of multiple attention heads, each learning distinct patterns of feature dependencies across the input sequence. By visualizing attention maps across these heads, we gain insights into how the model distributes its focus over the image.
Attention maps are particularly informative for ViTs because they directly represent the learned relationships between different image patches. In early layers, attention heads often capture local structures, while deeper layers aggregate global contextual information. By analyzing these attention distributions, we can interpret whether the model attends to semantically meaningful regions, such as object boundaries or discriminative features, aligning with human intuition for visual reasoning.

\subsection{Explainability of the models in classifying ceramic thin-section}

\begin{figure}[ht!]
    \centering
    \includegraphics[width=1\textwidth]{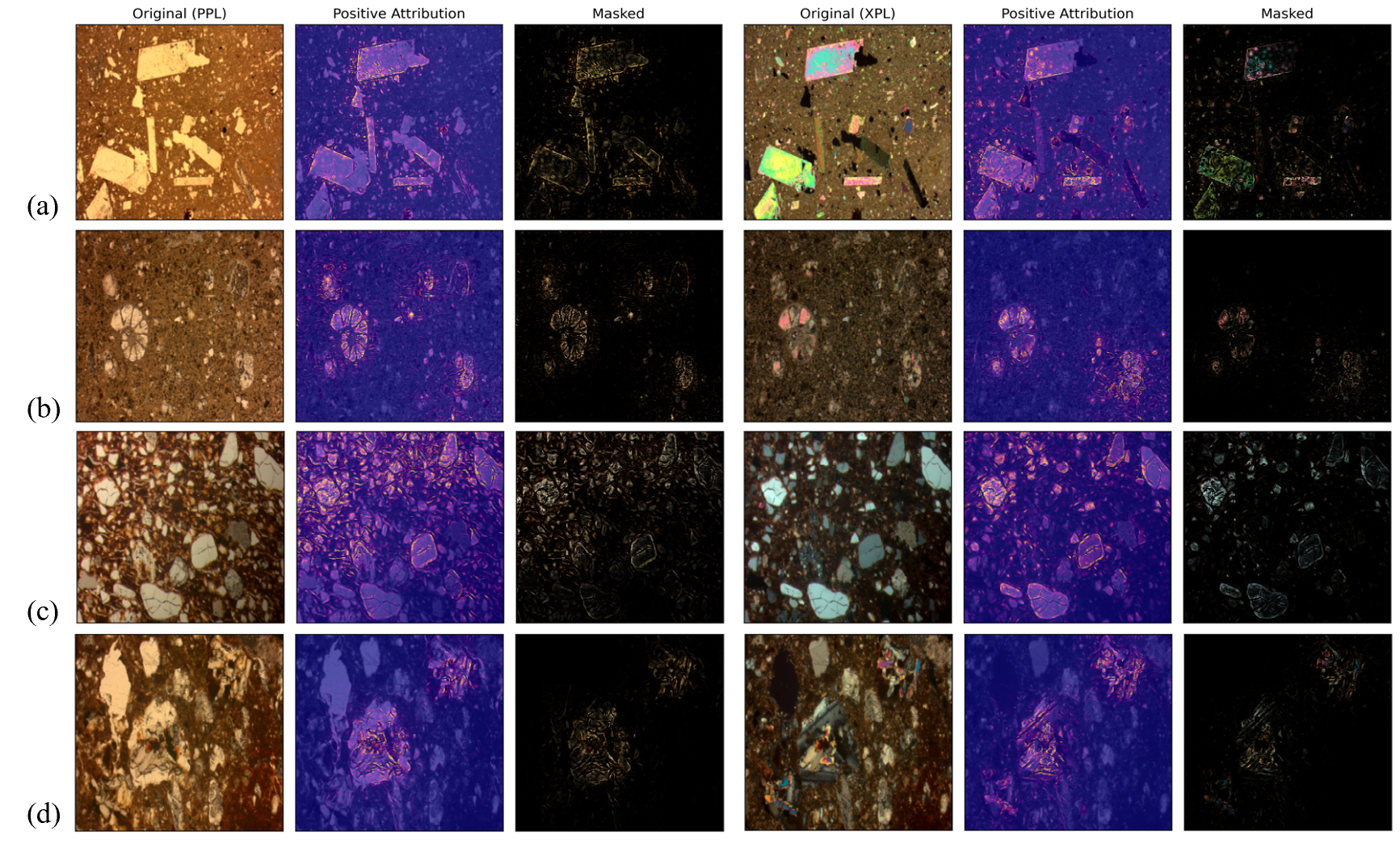}
    \caption{Guided Grad-CAM applied to pre-trained ResNet18. The figure shows the original images, the heatmaps overlapped to the thin-section images, and the masked images highlighting only the areas considered important for classification. (a) Class 1 (sample TF4): Calcite \textit{fabric}; (b) Class 2 (sample TF6): Carbonate A \textit{fabric}; (c) Class 6 (sample TFMcamp1): Quartz \textit{fabric}; (d) Class 8 (sample E16/2): Basalt \textit{fabric}.}
    \label{figure5}
\end{figure}

Guided Grad-CAM was used to visually explain the predictions of the pre-trained CNN model, highlighting the important features considered for classification. This technique allows us to observe how the model effectively identified the minerals characteristic of the \textit{fabrics} as representative for classification. The regions indicated by the Guided Grad-CAM as important correspond to the predominant minerals that lead to the discrimination and grouping of the ceramic samples according to their petrographic \textit{fabrics}. Fig. \ref{figure5} provides several significant examples, where the heatmaps allow for an immediate and clear visualization of the areas that positively contributed to the correct classification of the images. To analyze potential differences, we applied this method to both PPL and XPL images of the same samples. Identifying key features in PPL images is more challenging, as they appear less distinct compared to XPL images, making mineral discrimination more difficult. However, the results suggest that inclusions can still be distinguished from the matrix based on their shape and color. In XPL images, interference colors further help in distinguishing the main inclusions. Particularly interesting are the results for class 8. The images in Fig. \ref{figure5}d show how the heatmaps exclusively highlight the basalt fragments, distinguishing them from the other inclusions present in the thin section in both the PPL and XPL images. This demonstrates the neural network's ability to identify and learn specific characteristics of the corresponding petrographic \textit{fabrics}. See Fig. \ref{figureB1} in \ref{appendixB-XAI} to visualize the Guided Grad-CAM results for the other classes.

\begin{figure}[ht!]
    \noindent
    \centering
    \includegraphics[width=1\textwidth]{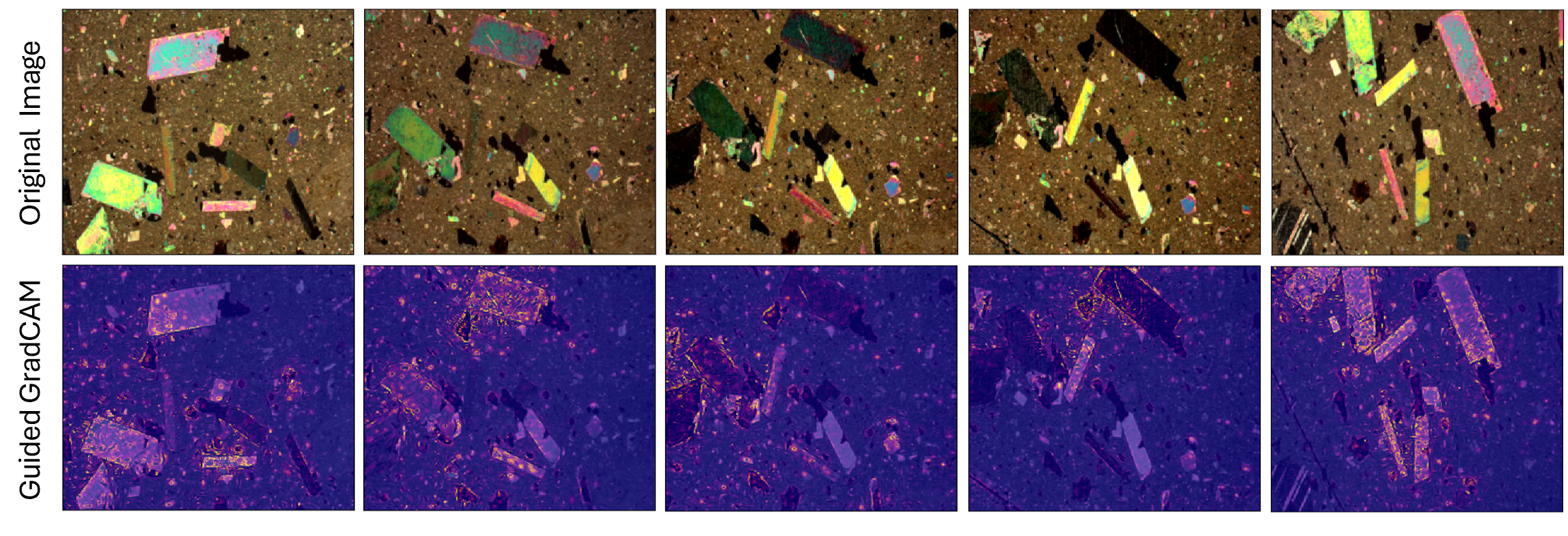}
    \caption{Guided Grad-CAM applied to the rotation sequence of sample TF4 from calcite \textit{fabric}.}
    \label{figure6}
\end{figure}

\begin{figure}[ht!]
    \noindent
    \centering
    \includegraphics[width=1\textwidth]{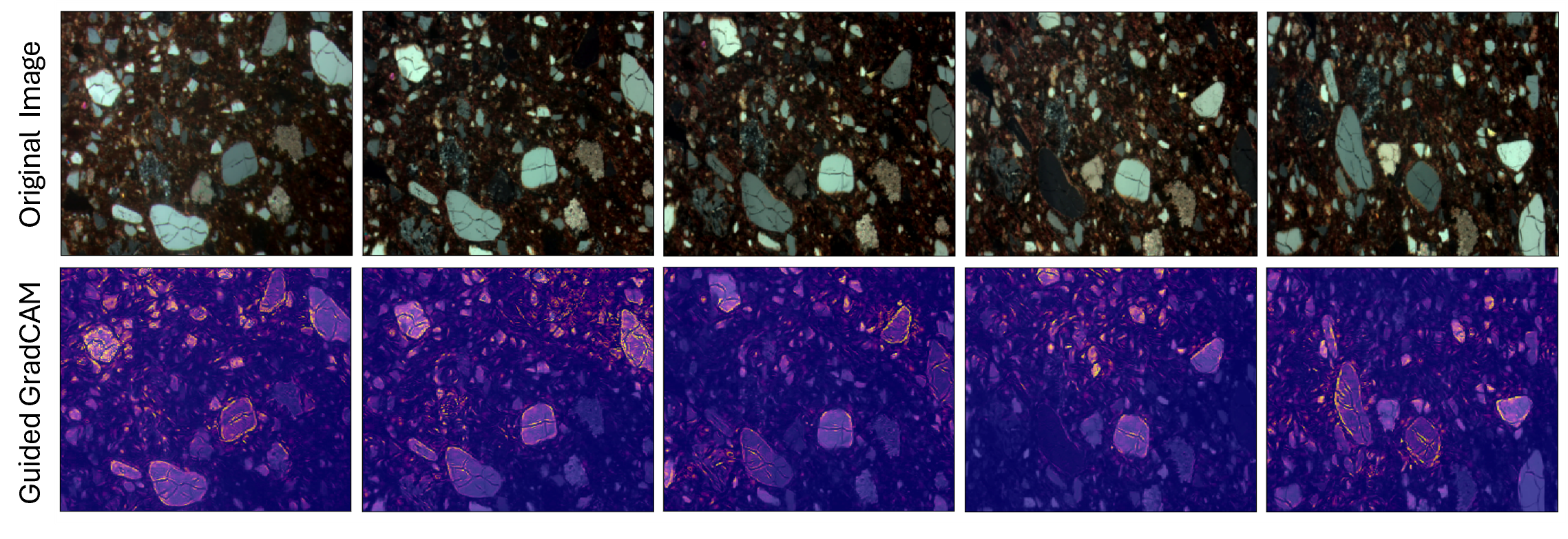}
    \caption{Guided Grad-CAM applied to the rotation sequence of sample TFMcamp1 from quartz \textit{fabric}.}
    \label{figure7}
\end{figure}

To deeper investigate these results, we applied Guided Grad-CAM to a sequence of images of the same sample at different rotations to determinate which parameters most influenced the classification and whether rotation and birefringence effects had any impact on the results. This analysis was conducted on calcite and quartz \textit{fabric} samples. Fig. \ref{figure6} shows the rotation sequence of a calcite \textit{fabric} sample, showing its progressive rotation until some crystals reached extinction. Calcite, characterized by high birefringence, is particularly distinctive in optical analyses and is a perfect case study to determine which features are considered important for classification. The resulting heatmaps indicate that rotation does not affect the recognition of specific patterns for classification. On the contrary, they highlight how the interference colors of the calcite crystals, their shape and edges, and also those in extinction are key elements considered by the model for classification. Regarding the analysis conducted on the representative quartz \textit{fabric} sample in Fig. \ref{figure7}, the heatmaps highlight quartz minerals independent of their size and/or brightness. Color appears to play a role in identification; in fact, the extincted quartz is not well evidenced, probably because the model is not able to discriminate it from the matrix as it is quite dark as well. Additionally, it is interesting to note that there are few carbonate rock fragments present in the sample that are not highlighted by the heatmaps. This result suggests that the model selectively recognizes quartz minerals without considering other inclusions, thus confirming the consistency of the classification process.  

\begin{figure}[ht!]
    \noindent
    \centering
    \includegraphics[width=1\textwidth]{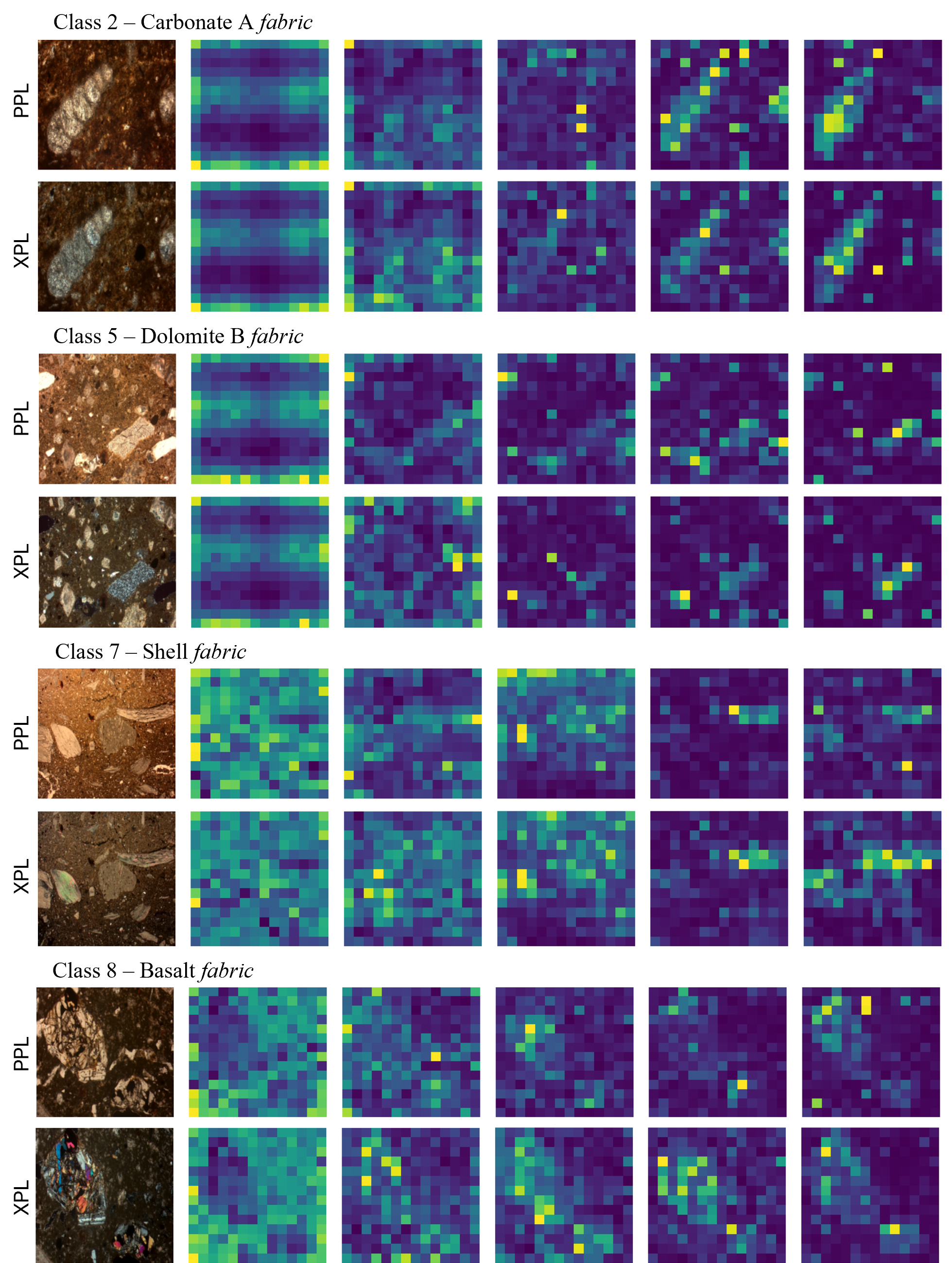}
    \caption{Attention maps of representative samples from carbonate A (sample KB146/30), dolomite B (sample K9), shell (sample KI11) and basalt (sample KB5) \textit{fabrics} both in PPL and XPL. Visualization of head 3 for class 2 and 5, and head 5 for class 7 and 8, across ViT’s layers.}
    \label{figure8}
\end{figure}

We use attention maps to visualize the classification predictions of the pre-trained ViT model, illustrating how it correlates different patches during the classification process. Fig. \ref{figure8} shows attention maps of representative samples from classes 2, 5, 7, and 8, which correspond to carbonate A, dolomite B, shell, and basalt \textit{fabrics}, respectively. Specifically, the figure presents attention maps from head 3 for class 2 and 5 and from head 5 for class 7 and 8 across different layers, allowing a better observation of the evolution of the attention over the layers of the ViT. To investigate potential differences in attention maps under different polarization conditions, we visualized the samples in both PPL and XPL. The results indicate that the attention maps remain consistent across both images, suggesting that the model's ability to focus on relevant mineralogical features is not significantly affected by polarization. Furthermore, the evolution of attention maps throughout the layers of the ViT appears consistent for each respective attention head. In the first layers, attention is more broadly distributed, while in deeper layers, it starts to focus on specific regions corresponding to the characteristic minerals of the petrographic \textit{fabrics}. 

The most challenging aspect of visualizing the important areas in this case is that the inclusions are distributed throughout the entire thin section. This makes it more difficult for the model to focus on specific features, especially considering that ViTs operate globally by analyzing patches rather than individual pixels. For this reason, it is particularly interesting to emphasize the results of classes 5 and 7, which contain a high number of inclusions dispersed in the matrix. In both cases, the attention maps show that the model was still able to focus on specific characteristics of the respective \textit{fabrics}. In class 5, the model mainly highlights some well-defined dolomite crystals that stand out clearly from the matrix, as well as the silicate rock fragment that is characteristic of this class. A similar pattern can be observed for the shell \textit{fabric}, where the attention maps seem to emphasize the areas corresponding to the elongated shell fragments rather than the more common carbonate fragments. We report the attention maps of representative samples from the other classes in \ref{appendixB-XAI}, Fig. \ref{FigureB2}.

Both the saliency and attention maps clearly highlight the characteristic minerals of the petrographic \textit{fabrics} as key features for classification. This indicates that the models distinguished the ceramic thin sections based on their inclusions, similar to the way traditional petrographic analysis groups ceramics by studying pores, inclusions, and the matrix. Although our models focused mainly on inclusions, it is interesting to note a certain correlation between deep learning-based classification and conventional petrographic methods.

\section{Conclusion}

This study demonstrated the potential of deep learning methodologies for the automated classification of ceramic thin sections based on their petrographic \textit{fabrics}, expanding on previous research by Lyons et al. \cite{Lyons2021, Lyons2022}. The application of Convolutional Neural Networks (CNNs) and Vision Transformers (ViTs) yielded high classification accuracy and enabled the grouping of thin sections with similar petrographic characteristics. Furthermore, explainability techniques, such as Guided Grad-CAM for CNNs and attention maps for ViTs, provided insights into the models' decision-making processes, confirming their ability to recognize key mineralogical features relevant for \textit{fabric} classification.

One possible limitation arises from the use of saliency maps, such as Guided Grad-CAM, which, while effective in highlighting important image regions, may not always provide a complete or precise explanation of model decisions. To address these challenges, future research could explore more recent explainability techniques, such as sparse autoencoders \cite{gao2024scaling}. Furthermore, integrating multimodal data sources, such as geochemical analysis alongside petrographic imaging, could improve classification accuracy and robustness, opening the possibility for more applications in the field of minero-petrographic studies. For instance, this approach could contribute to ceramic provenance research by aiding in the identification of raw material sources in relation to the geological context of surrounding areas.

A key future step is expanding the dataset of ceramic thin section images to improve model generalization and performance across a broader range of \textit{fabrics} and archaeological contexts. One current limitation is that models trained on a predefined set of known classes may struggle to generalize to previously unseen samples, particularly given the high degree of ceramic variability in the Levantine region, constraining their robustness in real-world applications. As a future step, it will be essential to evaluate the model on additional sets of thin-section images, distinct from those used during training. This will allow for a more realistic assessment of the generalization capabilities of the model, including its ability to identify samples that do not match any of the known classes (i.e., out-of-distribution detection). Moreover, recent advances in unsupervised learning and few-shot learning offer promising tools to address this challenge, potentially expanding the applicability of deep learning to archaeological scenarios.

Overall, this study highlights the potential of explainable AI in archaeometric research, providing a reproducible and efficient approach to ceramic analysis. Continued advancements in AI interpretability will be crucial for refining these methodologies, ensuring transparency, and facilitating broader adoption in archaeological and materials science applications.

\section*{Acknowledgements}

The authors acknowledge the support of Sapienza University of Rome. In particular, S. Scardapane and A. Devoto are partly funded by the Sapienza grant RG123188B3EF6A80 (CENTS). 

\appendix
\renewcommand{\thefigure}{\Alph{section}\arabic{figure}}
\renewcommand{\thetable}{\Alph{section}\arabic{table}}
\setcounter{figure}{0}
\setcounter{table}{0}

\section{Details on Model Architectures and Training}
\label{appendixA-hyperparams}

This section presents the architectural details and training configurations for both the ResNet-18 and DeiT-Small models used in our experiments. Hyperparameter selection was carried out based on the results of a 3-fold cross-validation performed on the training portion of the dataset (80\% of the full set). The best average validation results guided the final choice of hyperparameters. The explored hyperparameters are reported in tables A3 and A6. 

{\footnotesize
\begin{table}[H]
    \captionsetup{justification=raggedright, singlelinecheck=false}
    \caption{\label{tableA1} ResNet-18 Architecture Hyperparameters.} 
    \small
    \begin{tabular}{@{}l l c}
        \br
        \textbf{Parameter} & \textbf{Value} \\
        \mr
        Total layers & 18 \\
        Convolutional layers & 17 + 1 fully connected \\
        Residual blocks & 8 \\
        Input resolution & 224 × 224 \\
        Parameters & $\sim$ 11.7M \\
        \br
    \end{tabular}
\end{table}

\begin{table}[H]
    \captionsetup{justification=raggedright, singlelinecheck=false}
    \caption{\label{tableA2b} ResNet-18 Training Setup.}
    \small
    \begin{tabular}{@{}l l}
        \br
        \textbf{Hyperparameter} & \textbf{Value} \\
        \mr
        Learning rate & 3e-4 \\
        Weight decay & 3e-4 \\
        Training epochs (scratch) & 60 \\
        Training epochs (pre-trained) & 25 \\
        Batch size (scratch) & 20 \\
        Batch size (pre-trained) & 32 \\
        Optimizer & AdamW \\
        Learning rate scheduler & ReduceLROnPlateau \\
        \br
    \end{tabular}
\end{table}

\begin{table}[H]
    \captionsetup{justification=raggedright, singlelinecheck=false}
    \caption{\label{tableA5} ResNet18 hyperparameters explored during 3-fold cross-validation, for both pre-trained and from-scratch models.}
    \small
    \begin{tabular}{@{} l l }
        \br
        \textbf{Hyperparameter} & \textbf{Values Tested} \\
        \mr
        Learning rate & \{3e-4, 1e-4\} \\
        Weight decay & \{3e-4, 1e-4, 1e-5\} \\
        Optimizer & \{AdamW\} \\
        \br
    \end{tabular}
\end{table}

\begin{table}[H]
    \captionsetup{justification=raggedright, singlelinecheck=false}
    \caption{\label{tableA3} DeiT-Small Architecture Hyperparameters.}
    \small
    \begin{tabular}{@{}l l c}
        \br
        \textbf{Parameter} & \textbf{Value} \\
        \mr
        Layers (Transformer blocks) & 12 \\
        Heads & 6 \\
        Hidden dimension & 384 \\
        MLP dimension & 1536 \\
        Patch size & 16 × 16 \\
        Input resolution & 224 × 224 \\
        Parameters & $\sim$ 22M \\
        \br
    \end{tabular}
\end{table}

\begin{table}[H]
    \captionsetup{justification=raggedright, singlelinecheck=false}
    \caption{\label{tableA4b} Vision Transformer Training Setup.}
    \small
    \begin{tabular}{@{} l l }
        \br
        \textbf{Hyperparameter} & \textbf{Value} \\
        \mr
        Learning rate (from scratch) & 1e-4\\
        Learning rate (pre-trained) & 1e-5 \\
        Weight decay & 1e-4 \\
        Training epochs (scratch) & 125 \\
        Training epochs (pre-trained) & 60 \\
        Batch size & 32 \\
        Optimizer & AdamW \\
        Learning rate scheduler & CosineLR\\
        \br
    \end{tabular}
\end{table}
}

\begin{table}[H]
    \captionsetup{justification=raggedright, singlelinecheck=false}
    \caption{\label{tableA5} DeiT-Small hyperparameters explored during 3-fold cross-validation, for both pre-trained and from-scratch models.}
    \small
    \begin{tabular}{@{} l l }
        \br
        \textbf{Hyperparameter} & \textbf{Values Tested} \\
        \mr
        Learning rate & \{1e-4, 1e-5\} \\
        Weight decay & \{1e-4, 1e-5\} \\
        Optimizer & \{AdamW\} \\
        \br
    \end{tabular}
\end{table}

\section{Guided Grad-CAM and Attention Maps of Other Classes}
\label{appendixB-XAI}

In this section, we present the results of Guided Grad-CAM and attention maps for the remaining classes that were not shown in the main text of the paper.

\renewcommand{\thetable}{B\arabic{table}}
\renewcommand{\thefigure}{B\arabic{figure}}
\setcounter{table}{0} 
\setcounter{figure}{0}

\begin{figure}[ht!]
    \noindent
    \centering
    \includegraphics[width=1\textwidth]{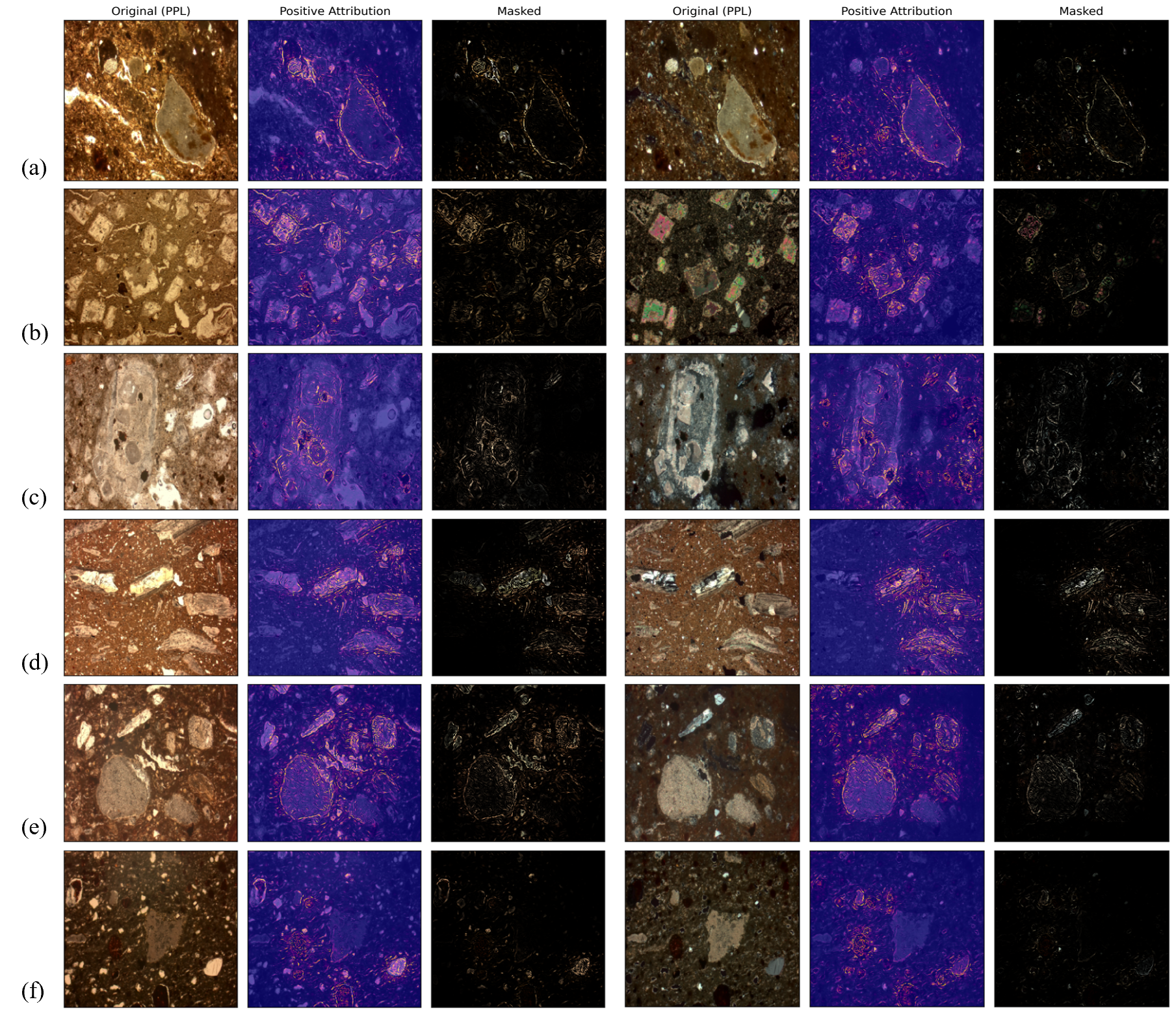}
    \caption{Guided Grad-CAM applied to pre-trained ResNet18. The figure shows the original images, the heatmaps overlapped to the thin-section images, and the masked images highlighting only the areas considered important for classification. (a) Class 3 (sample 718/25): Carbonate B \textit{fabric}; (b) Class 4 (sample B1/14): Dolomite A \textit{fabric}; (c) Class 5 (sample Kcamp15): Dolomite B \textit{fabric}; (d) Class 7 (sample KIcamp2): Shell \textit{fabric}; (e) Class 9 (sample K1): Carbonate C \textit{fabric}; (f) Classe 10 (sample KB383/38): Carbonate D \textit{fabric}.}
    \label{figureB1} 
\end{figure}

\begin{figure}[ht!]
    \noindent
    \centering
    \includegraphics[width=0.72\textwidth]{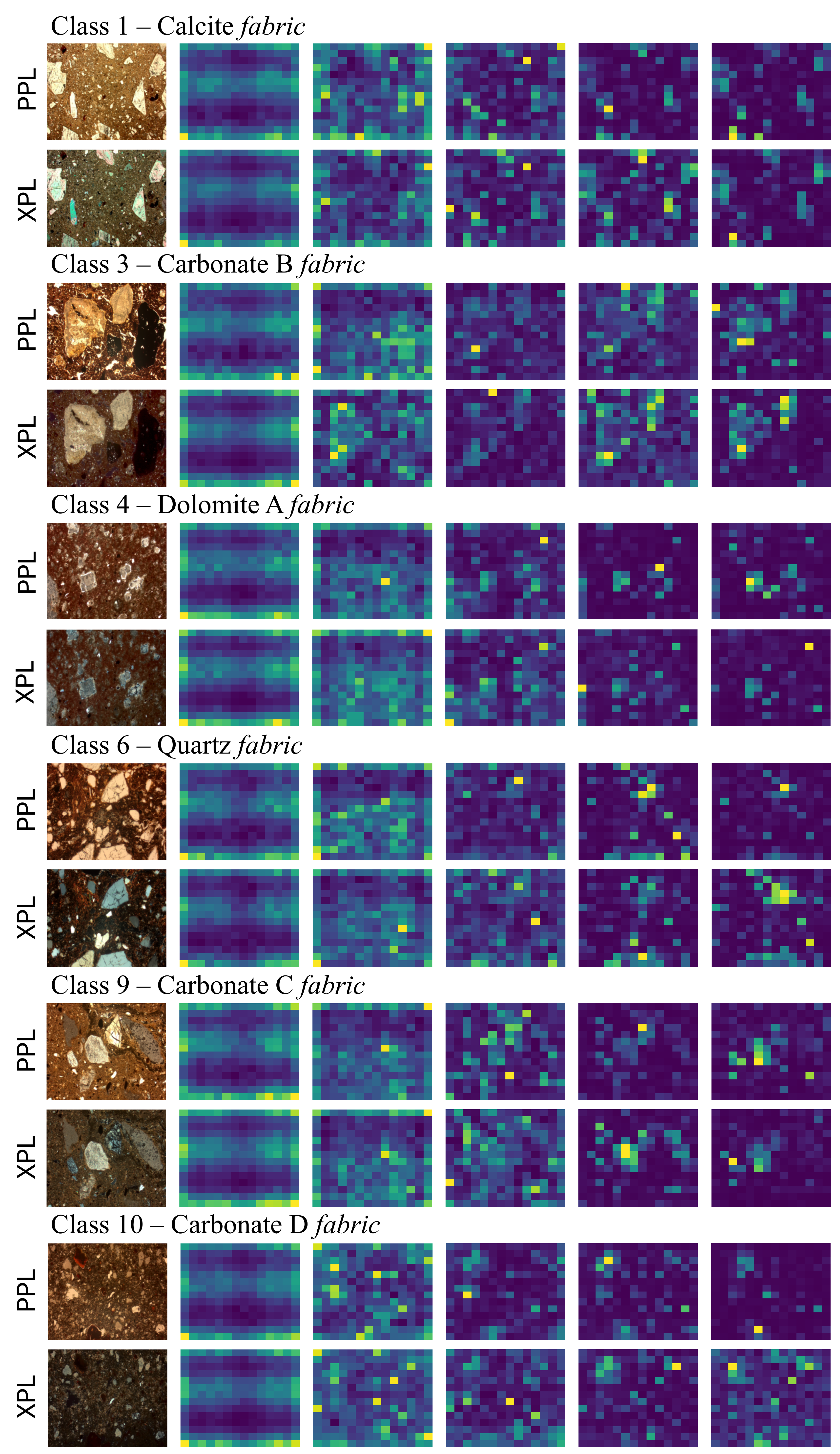}
    \caption{Attention maps of representative samples from calcite (sample TFcamp19), carbonate B (sample TFcamp13), dolomite A (sample Asur5), quartz (sample KBC4), carbonate C (sample K16) and carbonate D (sample KB314/113) \textit{fabrics} both in PPL and XPL. Visualization of head 3 across the ViT's layers.}
    \label{FigureB2}
\end{figure}

\clearpage
\section*{References}
\bibliographystyle{unsrt} 
\bibliography{bibliography}

\end{document}